# Geologic Disposal Safety Assessment (GDSA) Biosphere Model Development

Caitlin Condon [a] (*caitlin.condon@pnnl.gov*), Saikat Ghosh [a], Bruce Napier [a], Glenn Hammond [a], Harish Gadey [a]
[a]*Pacific Northwest National Laboratory, 902 Battelle Blvd, Richland, WA 99354*

## INTRODUCTION

The Spent Fuel and Waste Science and Technology (SFWST) Campaign of the U.S. Department of Energy Office of Nuclear Energy (DOE-NE), Office of Spent Fuel and Waste Disposition is conducting research and development (R&D) on geologic disposal of spent nuclear fuel (SNF) and high-level nuclear waste (HLW). This work includes the Geologic Disposal Safety Assessment (GDSA) program which is charged with development of generic deep geologic repository concepts and system performance assessment models. One part of the GDSA framework is the development of a biosphere model capable of assessing dose to potential receptors exposed to radionuclides released via groundwater from geologic disposal sites. This work includes the development of a biosphere model capable of estimating doses to potential receptors living in the biosphere and exposed to radionuclides released from a hypothetical geologic disposal site.

The GDSA biosphere model is being developed so that it is compatible with the GDSA framework including PFLOTRAN, the massively parallel subsurface flow and reactive transport code. PFLOTRAN is a subsurface flow and reactive transport code that solves a system of partial differential equations for multiphase flow and transport of components in porous materials such as shale formation [1]. PFLOTRAN simultaneously simulates energy and mass flow with fluid properties as function of pressure and temperature through equations of state. PFLOTRAN also solves the mass conservation and transport equations of multicomponent formulations of aqueous chemical species, gases, and minerals reactive transport. It contains a waste form process model that simulates the radionuclide inventory under potential failures in a geological repository. PFLOTRAN can thus calculate the source term of radionuclides in ground water which can then be used as the GDSA biosphere model input.

The purpose of the GDSA biosphere model is to be a flexible, open-source code that will be compatible with the PFLOTRAN subsurface model. PFLOTRAN groundwater radionuclide concentrations are used as input in the GDSA biosphere model to calculate dose to a hypothetical reasonably maximally exposed individual living in the biosphere. The biosphere model was designed to be a generic assessment tool that can be used to assess the dose and risk associated with a wide variety of sites and climates. The design considered the potential future use cases and stakeholder needs regarding a repository biosphere model and the design's flexibility is intended to meet these stakeholder needs.

The GDSA Biosphere model was designed to be consistent with international recommendations and guidance for environmental assessment of deep geological repositories. The GDSA biosphere model design relies on information and guidance from the International Atomic Energy Agency (IAEA) including the BIOMASS program as well as guidance from BIOPROTA [2-4]. The Organization for Economic Co-Operation and Development Nuclear Energy Agency (NEA) has developed and published a list of FEPs that can help evaluate the "long-term safety or performance" of a geologic repository [5]. This list of FEPs was used as a guidance document when determining which pathways of exposure and variables needed to be incorporated in the GDSA Biosphere model.

This work summarizes the biosphere model development including the interfacing with the PFLOTRAN subsurface model, setting up the GDSA biosphere scenarios, and the models used to calculate radionuclide transport through various environmental media.

## MODEL THEORY

The GDSA framework's biosphere model is a biosphere model compatible with the PFLOTRAN subsurface model. PFLOTRAN simulates the deep groundwater transport of radionuclides released from a deep geological repository to the biosphere. Radionuclides enter the biosphere system through the local groundwater aquifer from a hypothetical release from a geologic repository (Fig. 1). The biosphere model calculates the specific activity from these input concentrations based on the user-established scenario such as whether the user chooses groundwater or surface water as the source term for domestic water supply, agricultural water supply, or recreational surface water use. Contamination of a surface water body from an interconnected aquifer is calculated using a user provided dilution factor. A primary requirement of the GDSA biosphere model is to track radionuclide decay and ingrowth in environmental media. The GDSA biosphere model tracks each radionuclide individually through the user-selected exposure pathways while tracking the ingrowth of each decay product individually through all applicable environmental media (e.g., crop, soil, surface water). The GDSA biosphere model also captures decay of radionuclides in environmental media over time.

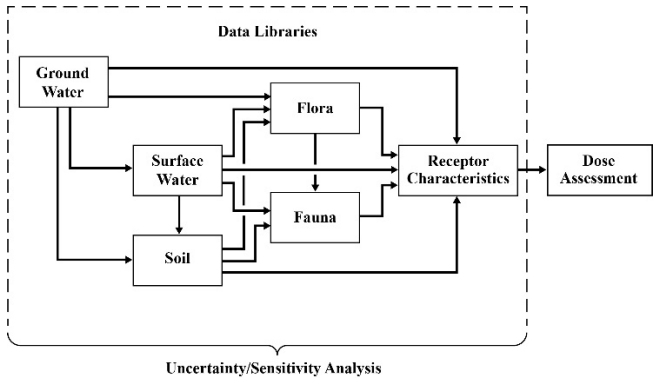

Fig. 1. GDSA Biosphere Conceptual Framework

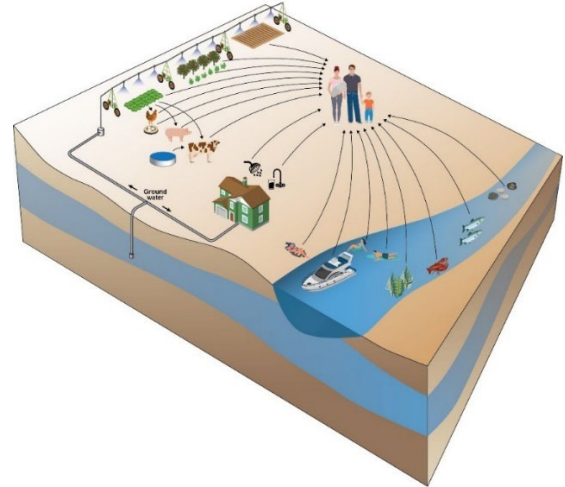

Fig. 2. Biosphere Model Exposure Pathways

The radionuclides from the ground water are introduced to the biosphere environmental media. For example, the radionuclides introduced into the GDSA biosphere model may result from irrigation deposition and are partitioned among various media compartments assuming equilibrium steady state conditions. The suite of possible doses to humans are calculated for various exposure pathways such as ingestion of radionuclides in drinking water, terrestrial food (leafy vegetables, root vegetables, fruits and cereals), animal food (milk, meat, egg), aquatic food (fish, mollusks, crustaceans, and aquatic plants); inadvertent ingestion of contaminated soil; inhalation of resuspended soil dust; ingestion of surface water during swimming; external radiation from radionuclides in residential soil, river beach and swimming water (Fig. 2). Libraries of radionuclide specific transfer factors are used to model equilibrium transfer between various media components such as plant crops, animals, and aquatic life. Exposure pathways are set up based on present-day human behavior and activity based on factors provided in the *Exposure Factors Handbook* [6]. As currently configured, dose coefficients are based on ICRP 60 and risk factors are based on Federal Guidance 13. The database of radionuclide specific dose and risk coefficients were extracted for the adult age group from Dose Coefficients Package (DCFPAK) software [7].

The GDSA biosphere model tracks the radionuclide decay and progeny ingrowth as well as radionuclide transfer between media compartments to support assessment of exposure to the human receptors. During environment transport from water to the various transfer media of the biosphere to humans, a radionuclide may undergo decay and progeny radionuclides may be introduced (dependent on the decay physics of the radionuclides included in the source term). Scenario dose and risk assessment account for the radiological emissions from both the radionuclides included in the initial source term from the repository and any progeny that may be introduced to the biosphere as a result of the decay of the initial source term as releases move through the biosphere. The decay and ingrowth of radionuclides is modeled using a system of linear first order differential equations that accounts for branching and loss due to removal from a media compartment. In this form, the radionuclides can be removed from a media compartment (such as soil) either through physical loss (such as leaching) or transfer to another compartment (such as plant uptake), or the radionuclides can decay, in place, and be transformed into another radionuclide [8]. The generalized differential equation for the rate of change of a radionuclide in a given compartment is shown in eq.1:

$$\frac{dA_c(t)}{dt} = -L_c A_c(t) - \lambda_{rc} A_c(t) + \sum_{n=1}^{c-1} d_{nc} \lambda_{rn} A_n(t)$$
(1)

Where,

$L_c$ = total rate constant for all physical transfers of chain member from the medium (days$^{-1}$),
$A_c(t)$ = quantity of chain member c at time t (atoms),
$\lambda_{rc}$ = radioactive-transition rate constant for chain member c (days$^{-1}$),
$A_n(t)$ = quantity of chain member n at time t (atoms),

$d_{nc}$ = fraction of precursor radionuclide transitions (chain member n) that result in production of the chain member c (dimensionless), and

$\lambda_{rn}$ = radioactive-transition rate constant for chain member n (days$^{-1}$).

A general solution to these differential equations was provided by Kennedy and Strenge and is integrated within the GDSA biosphere model [8]. The general solution calculates the time-integrated concentrations of each radionuclide in a certain media.

**MODEL DEVELOPMENT & PROGRESS**

The GDSA biosphere model is being developed using an object-oriented design in Fortran 90 with extensible features for Fortran 2003. Objects contain the data structures and procedures necessary to implement a specific function, and interfaces are set up for interaction between objects. The biosphere model incorporates the coding style and some features from the PFLOTRAN model. Object-oriented design allows flexibility for future modifications to an object's data structures and functions with little or no impact on other portions of the code [10]. The biosphere model code is being developed as a nested hierarchy of dynamically allocated objects from the highest-level simulation object down to low-level auxiliary objects that store all state variables for each exposure pathway. Given a pointer to the top-level simulation object, the developer has access to all underlying data structures or objects. This hierarchy engenders modularity and structure within the code.

The code takes inputs for a specific biosphere scenario from a user provided text file which contains input data for various exposure pathways and environmental parameters. The model is flexible in order to simulate either a single or multiple exposure pathways based on user selections in the input file. Radionuclide activity for the source term is calculated from an input file of time-dependent radionuclide concentrations from PFLOTRAN simulation. A python script directly formats the PFLOTRAN output for input to the biosphere model containing groundwater concentration data of one or more radionuclides as a function of time in years. Though users may simulate PFLOTRAN for up to million years, the GDSA biosphere code allows users to select a subset of such simulation time periods. For example, a user may choose 100,000 years as the starting point for human exposure and an exposure period of 70 years for an average individual. Thus, the GDSA biosphere code will extract and interpolate the radionuclides' concentrations over the period of 100,000 to 100,070 years and subsequently will calculate the dose for each year in the exposure time period. The biosphere model will be flexible enough to allow for accumulation from irrigation deposition prior to start of exposure (e.g., irrigation for 10 years prior to beginning of dose estimates). The accumulated radionuclides in environmental media will carry over to the next year when the exposure to humans is simulated. The ability to set up different GDSA biosphere scenarios at different time points of interest within a PFLOTRAN simulation allows users to capture potential climate state changes at a given location.

As stated previously, the biosphere code is designed to set up the decay chain for each radionuclide within the PFLOTRAN simulation output. The model is therefore capable of calculating dose and risk to humans from exposure to multiple radionuclides through one or more pathways.

The decay chain is set up based on a radionuclide data library that contains all associated radioactive decay data including half-lives, decay chains, and branching ratios. The radionuclides are organized into decay chains ordered by atomic number under the radionuclides highest in the chain. Some "implicit" progenies are included in few instances without the indication of a half-life. These "implicit" progeny have very short half-lives (<10 minutes) and their transient nature does not affect the activity of other radionuclides in the series. However, they are included in the file to account for their energy emissions during dose calculations. Transfer factors, human behavioral activity, and dose and risk factors are stored in text-based databases from which values are extracted for each radionuclide in a specific decay chain during simulation.

Uncertainty simulations with the GDSA biosphere code are configured and set up with the GDSA framework Dakota software [11]. The Dakota software, originally developed at SNL, contains state-of-the-art algorithms for optimization, uncertainty and sensitivity analysis, and parameter estimation. A user does not require detailed knowledge about the underlying software package in Dakota. The GDSA biosphere code is being developed to implement the Monte Carlo based uncertainty characterization of the biosphere model to the variability in user parameters, transfer factors, and PFLOTRAN concentrations. The GDSA biosphere model will use Dakota to perform uncertainty analysis on biosphere scenario runs (Fig. 1).

A prototype of the GDSA biosphere model has been developed and tested with one full pathway of exposure for human receptors. The prototype case starts with a PFLOTRAN groundwater source term to mimic well water (or to be diluted to surface water); that water is used to irrigate terrestrial crops, those crops are harvested, and then those crops are consumed by the human receptors. Currently, the first full iteration of the GDSA biosphere model with all exposure pathways is under development.

**CONCLUSIONS AND FUTURE WORK**

The GDSA biosphere model is a biosphere model component of the GDSA framework. The GDSA biosphere model compatibly uses the PFLOTRAN subsurface model output. The purpose of the GDSA biosphere model is to be a flexible, open-source code that can be used to determine the dose to a hypothetical maximally exposed individual living in the biosphere given a PFLOTRAN groundwater radionuclide concentration model input. The GDSA biosphere model prototype developed in fiscal year 2021 included one full pathway from groundwater to a human receptor dose. The full GDSA biosphere model is currently under development to include all remaining exposure pathways identified in this paper as well as developing a test case scenario for model testing and applying uncertainty using Dakota. PNNL-SA-174306.